\documentstyle[multicol,prl,aps,epsf,psfig,epsfig]{revtex}
 
\begin{document}
  
 \title
 {
Persistence and dynamics in ANNNI chain
 }
 \author
 {
	  Parongama Sen and Subinay Dasgupta
  }
\address
  {
	  Department of Physics, University of Calcutta,
	      92 Acharya Prafulla Chandra Road, Kolkata 700009, India. \\
	      parongama@vsnl.net,subinay@cubmb.ernet.in\\
  }
\maketitle
\begin{abstract}
We investigate both the local and global persistence
behaviour in  ANNNI (axial next-nearest neighour Ising) model.
 We find that
when the ratio $\kappa $ of the second neighbour interaction 
to the first neighbour interaction is less than 1,
$P(t)$, the probability of a spin to  remain in its original state
upto time $t$ shows a stretched exponential decay.
For $\kappa > 1$, $P(t)$ has a algebraic decay but the exponent is different
from that of the  nearest neighbour Ising model.
The global persistence behaviour  shows similar features.
We also conduct some deeper investigations in  the dynamics of
the ANNNI model and conclude that it has a different dynamical behaviour 
compared to the nearest neighbour Ising model.


Preprint number: CU-Physics-01/2004
\vskip 0.5cm

\end{abstract}
\begin{multicols}{2}
\bigskip
\section{Introduction}

The tendency of  a spin in an Ising system to remain in its original
state following a quench to zero temperature has been extensively studied
over the last few years and is well-established as an example of the
phenomenon called persistence in dynamical systems 
\cite{bray}. Quantitatively, persistence is measured 
by the probability $P(t)$ that a spin does not flip upto time $t$. 
$P(t)$  shows a power law behaviour, i.e.,  $P(t) \sim t ^{-\theta}$, 
where $\theta$ is a new exponent not related to any other 
static or dynamic exponent. This phenomenon has been observed
and studied quite extensively in Ising models with nearest neighbour 
interaction  in different dimensions \cite{bray,hakim,stauffer}.

Apart from such `local' persistence, one can also study the `global' 
persistence \cite{satya_globe} behaviour by measuring the probability 
$P_G(t) $ that the order parameter does not change its sign till time $t$. 
At the critical temperature, the probability that the individual spins will
not be flipped till time $t$  has an exponential decay, while the 
global persistence shows an algebraic decay: $P_G(t) \sim t^{-\theta_G}$.
The critical temperature of the Ising model in one dimension
being zero, the global persistence becomes a quantity of interest together 
with the local persistence at zero temperature.
Exact values
of the exponents for the nearest neighbour Ising chain 
are known to be $\theta=0.375$ \cite{hakim}, and $\theta_G = 0.25$ \cite{satya_globe}.

In this paper, our objective is to investigate the effect of frustration on
(local and global) persistence, by computer simulation of an ANNNI 
(axial next-nearest neighour Ising) chain 
\cite{selke} at zero temperature. This model has the Hamiltonian
\begin{equation}
H = -\Sigma_{i=1}^L (S_iS_{i+1} - \kappa S_i S_{i+2}),
\end{equation}
where $S_i$ is the spin ($\pm 1$) at $i$-th site and $\kappa$ ($>0$) is the
parameter which represents the amount of frustration. We choose this 
particular model for our study because it is perhaps the simplest classical
model with tunable frustration and because this model as such shows very 
interesting
static and dynamic behaviour \cite{selke,barma,redner,chang}. As regards static
behaviour, the ground state is ferromagnetic for $\kappa < 0.5 $, antiphase
(++ - - type) for $\kappa > 0.5$ and highly degenerate for $\kappa = 0.5$. 
On the other hand, zero-temperature dynamics using single spin-flip does
not lead to the ground state for $0 < \kappa < 1$ but does lead to the ground
state for $\kappa > 1$ \cite{redner}.

Our main  observation is that for $0< \kappa < 1$  there is no persistence 
(i.e., no algebraic decay of $P(t)$)
while for $\kappa > 1$ there {\em is} persistence 
albeit with a persistence exponent different from that of the unfrustrated
nearest neighbour Ising model (i.e., $\kappa = 0$).
We also claim, using some novel approaches,  that as regards domain dynamics, the ANNNI model belongs to a 
different dynamical universality class for $\kappa>1$.

In section II, we describe the model and the studies on local persistence for
$\kappa < 1$ and $\kappa > 1$.
In section III, the  dynamics of the domains for $\kappa > 1$ 
is analysed in detail.
The global persistence behaviour is presented in section IV. Since the behaviour at $\kappa=1$ is
unique, we have discussed it in a separate section (section V).
The results are summarised and  discussed  in section VI.

\section {Local persistence in ANNNI chain}

{\it The Model} 

We take an ANNNI chain of $L$ spins in one dimension with periodic boundary 
condition in a random 
configuration (infinite temperature) and quench it to zero temperature. Thus, 
our updating rule is that a spin is selected randomly from the system,
it is flipped (not flipped) if its energy is positive (negative), and
it is flipped with probability 0.5 if its energy is zero.
$L$ is always chosen to be a multiple of 4 to ensure complete
antiphase ordering. We have used $L=$ 8000 - 12000 (unless otherwise mentioned)
and averaged the results over $10^3$ to $10^4$  configurations as per necessity.

\pagebreak

Case I : $\kappa < 1$

For local persistence, in the entire range $ 0 < \kappa < 1$
the decay of $P(t)$  is not algebraic (Fig. 1), but rather a stretched exponential,
\begin{equation}
P(t) \sim \exp(-\alpha t^\beta)
\end{equation}
with $\alpha = 1.06$ and $\beta = 0.45$. The exponent $\beta$ as
well as the coefficient $\alpha$ is found to 
be independent of $\kappa$. There is no special behaviour 
at $\kappa = 0.5$, the multiphase point for the ground state. 

\begin{figure}
\noindent \includegraphics[clip,width= 6cm, angle = 270]{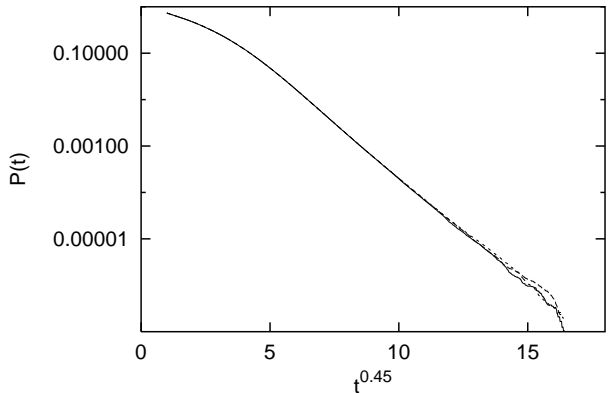}
\caption{The variation of $P(t)$  for $\kappa <1$ are shown for 
 $\kappa$ = 0.2,0.4,0.6 against $t^{0.45}$. The curves fit fairly well
 to a linear form in the log-linear plot indicating that $P(t)$ fits to a 
  stretched exponential form  $P(t) \sim \exp(-1.06t^{0.45})$.
 The local persistence seems to be independent
of $\kappa$. Here the simulations have been done
for $L=12000$.}
 \end{figure}
The dynamical processes occurring in the system are as follows. In the first few
dynamical steps ( - the precise number depends on the system size) domains of 
size 1 are removed and one is left with domains
of size $\geq  2$. No domain wall is annihilated or created henceforth but 
the dynamics continues indefinitely as all  domains of size 
$ > 2$ are unstable. As a result, the system attains a  non-equilibrium 
steady state but the equilibrium state is never reached and
in a finite time all the spins in the system are 
flipped. This justifies the faster than   power-law decay of $P(t)$ in 
the system. To illustrate our argument further, we observe from simulation
studies that after the first few iterations (typically 10 iterations for a 
system of $L=16000$ spins) the number of domain walls per spin (say $M/L$) attains a 
constant value of $0.2795$.  This constant value of $M/L$ is quite close 
to the most probable value of $M/L$ which can be theoretically  estimated easily.
Under the constraint that each domain is of size $ \geq 2$, the number of 
configurations in a system of $L$ spins and $M$ domains, is
\begin{equation}
      N(L,M) = {{L-M} \choose {M}},
\end{equation}
and the maximum value of this quantity is for
\begin{equation}
   \frac{M}{L} = \frac{(\sqrt{5}-1)}{2\sqrt{5}}=0.2764.
\end{equation}

Case II : $\kappa > 1$

For local persistence, in the range $ \kappa > 1$, 
the behaviour of $P(t)$  agrees well with an algebraic decay
\begin{equation}
P(t) \sim  t^{-\theta^{\prime}}
\end{equation}
at large $t$ (Fig. 2).
However, $\theta^{\prime}$ shows a weak dependence on the
time interval over which it is calculated indicating that there is 
a correction to the scaling. We have  calculated 
the slopes in the log-log plot  over different intervals of time
and found that
$\theta ^{\prime}$ increases slowly with time 
initially but
at large times ($t > 10000$) attains convergence
to $\theta^{\prime} =
  0.69 \pm 0.01$. This value is considerably
different from that of the ferrromagnetic Ising model. 
The value of ${\theta}^{\prime}$ remains the same for all $\kappa$ except at the
point $1/\kappa = 0$, where the system breaks up into two independent 
sublattices with nearest-neighbour antiferromagnetic interactions,
and ${\theta}^{\prime}$ becomes equal to $\theta$. The dynamics here
is such that the patches of antiphase state (e.g., ++ - - ) grow in 
size gradually and the equilibrium (lowest-energy) configuration of 
antiphase is reached eventually. Similar dynamics also prevails in the 
ferromagnetic state ($\kappa = 0$), as the regions of up (or down) spins grow
in size and ultimately fill up the entire system. However, one must note that
inspite of the similarity in {\em dynamics}, the {\em persistence exponents} 
are indeed different. In the next section we investigate the domain dynamics 
for $\kappa > 1$ in greater detail.
\begin{figure}
\noindent \includegraphics[clip,width= 6cm, angle = 270]{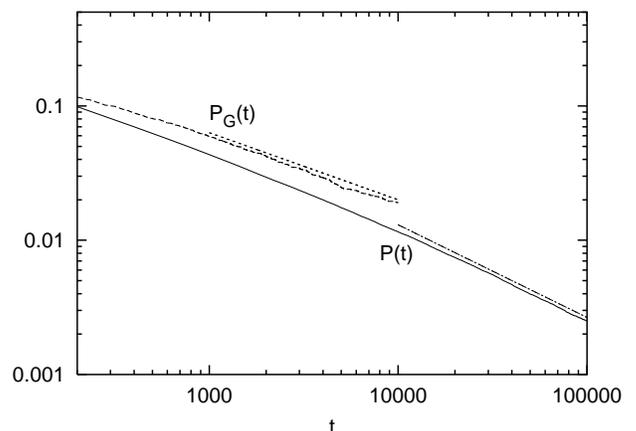}
\caption{The variation of $P(t)$ and $P_G(t)$ are shown for $\kappa$ = 2.0.
 Both have a power law decay, the best fit straightlines in the log-log
 plot for large $t$ show that $P(t)$ has an exponent $\sim 0.69$ while $P_G(t)$ has an
exponent $\sim 0.50$ (see text for the details in calculating the exponent $\theta ^\prime$). 
The curves are the same for all $\kappa > 1$. Here the simulated system sizes are  
$L=10000$  and $L=8000$ for $P(t)$ and
$P_G(t)$ respectively.}
 \end{figure}

\section{Domain dynamics for $\kappa > 1$}

In some earlier studies \cite{barma,redner}, 
it has been
concluded that the ANNNI model belongs to a different dynamical universality 
class  compared to the nearest neighbour Ising model. 
However, precise values of exponents for other dynamical processes e.g., 
dynamical exponent, are not available for the ANNNI model. 
In the nearest neighbour ferromagnetic Ising case, the dynamical exponent $z=2$.
Here we have attempted to estimate the value of $z^\prime$, the dynamic exponent for
the ANNNI chain.

To estimate $z ^\prime$ directly,
we have  calculated  the lifetimes $\tau$ defined
as  the time upto which the dynamics continues in the ANNNI model as a
function of the system size $L$ (Fig. 3). In the nearest-neighbour Ising model,
this time varies as $L^{z}$, while for the ANNNI model we find
\[
\tau \sim L^{z'},
\]
with $z^\prime \simeq 2.15$.
\begin{figure}
\noindent \includegraphics[clip,width= 6cm, angle = 270]{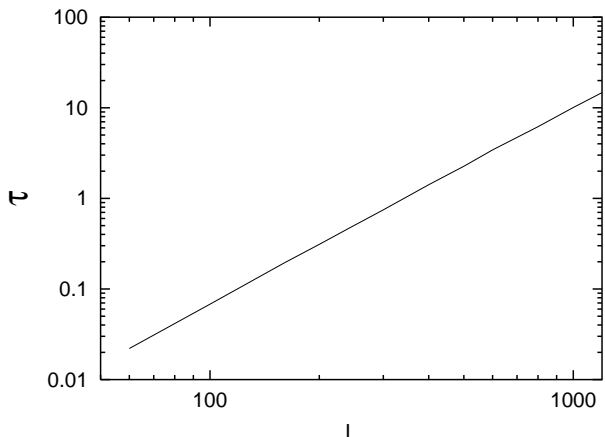}
\caption{
Lifetime $\tau$ (scaled down by $10^5$)  is shown against the system size $L$.
 $\tau$ increases as $L^{2.15}$. 
The value of $\kappa = 2.5$ here.}
 \end{figure}

The dynamics in the nearest-neighbour Ising model can be viewed as a random 
walk  of the domain walls which separate regions of up-spins and down-spins. 
The domain walls annihilate each other when they meet and their
number $M$ reduces with time as  $t^{-1/z}$ with the value of $z$  
equal to $2$ as mentioned above. 
On the other hand, in the ANNNI model, the domain dynamics is different and the
domain walls have  a different role in the antiphase state.
In order to compare the dynamics in these two systems, it is therefore
useful to regard the domain walls in the non-frustrated ($\kappa=0$) Ising 
model as ``defects'' and compare its dynamics with that of an analogous 
quantity in the ANNNI model.  We have adopted a number of ways to
estimate the defects in the ANNNI model (for $\kappa > 1$)
as a function of time to get an estimate of the domain decay  exponent.   
In the following we briefly describe these methods.

First we note that the excitation energy $\Delta E(t)$ (deviation from the 
ground state energy) of any state in the nearest neighbour ferromagnetic 
Ising model is identical to the number of domains (apart from some 
multiplicative constant), and as the  system relaxes to its equilibrium 
ground state, $\Delta E$  also shows a decay $\Delta E \sim t^{-1/2}$.    
In the ANNNI model also, we study the relaxation of energy by computing
$\Delta E$ which shows the behaviour (Fig. 4)
$\Delta E \sim t^{-1/z'} $ with $z' \simeq 2.40$. We claim that $z'$ is 
the dynamic exponent in the ANNNI model.

\begin{figure}
\noindent \includegraphics[clip,width= 6cm, angle = 270]{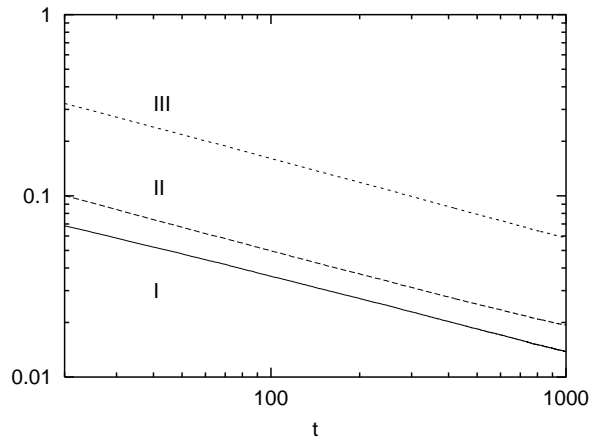}
\caption{The time decay of $M '$  (number of domains of length
other than 2; curve maked I),
of ``defect'' sites $D$ (II) and of excitation energy $\Delta E$ (III) 
are shown against time $t$ ($\kappa = 2.5$). All the simulations have been done
for  systems with $L=1000$ spins.} 
 \end{figure}

Secondly, a direct measure of the defects defined to be simply the
spins that do not belong to ++ - - type states has also been done 
(e.g., in the distribution ++ - - +++ - -, the 7-th spin is a defect, while
in ++ - - ++++ the 7-th and 8-th spins are defects.) Our simulation
shows that the number of such defects, $D$, {\em also} decreases as 
$t^{-1/z'}$  for all $\kappa > 1$ with $z^\prime \simeq 2.30$.

A third way to estimate $z'$ is to count the number of domains which are
not of length 2, i.e., domains which do not satisfy the antiphase
configuration. As mentioned earlier, in \cite{redner},
the dynamics of domains of different lengths were studied from which it was 
concluded that the ANNNI chain belongs to a new universality class.
Here we count all domains of size other 
than two. We find a power law variation of this quantity (say $M'$) again 
with exponent $1/z'$ where $z' \simeq 2.30$. 
All the above  results are shown in Fig. 4. 

From all the above estimates, we  conclude that the dynamic exponent exponent 
is $z' = 2.3 \pm 0.1$ for 
the ANNNI model in one dimension.

It is possible, in principle, to calculate the value of $z^\prime$ also from the 
relation
\begin{equation}
P(t \to \infty,L) \sim L^{-z^\prime \theta^\prime}, 
\end{equation}
when the dynamics has stopped altogether \cite{puru}. However, such an estimate would require
simulations upto $L^{z^\prime} $ number of Monte Carlo steps for a system of size 
$L$. We could get these data only for systems with $L \leq 1000$ which is insufficient
to give a good scaling of $P(t \to \infty,L)$ with $L$.

\section{Global persistence}

We shall now present our results on {\em global} persistence. In this
context one needs to be careful in defining the order parameter
for different values of $\kappa$. For $\kappa < 0.5$, the global persistence
may be calculated in terms of the magnetisation as for the ferromagnetic
nearest-neighbour Ising model and the global persistence shows an
exponential decay :
\begin{equation}
P_G(t) \sim \exp (-0.10 t).
\end{equation}
In the entire range $0.5 < \kappa < \infty$, one needs to define the antiphase
order parameter. It is not easy to put forward such a definition in a
straightforward manner. One may be tempted to define the sublattice
magnetisation

\[
S_1 + S_2 - S_3 - S_4 + S_5 + S_6 - S_7 - S_8 + \cdots
\]
as the order parameter but such a definition does not work because the
antiphase has a four-fold degeneracy. We propose instead to consider the
magnetisations over four sublattices : \\
\begin{equation}
m_\alpha = \Sigma_{j=0}^{L/4-1} S_{\alpha + 4j}; ~~~\alpha =1,2,3,4
\end{equation}
as the order parameter. The global persistence function $P_G(t)$ is evaluated
for each sublattice and an average over the four sublattices is taken. 
We have verified that for $\kappa = 0$, with this definition of order 
parameter one gets a  power-law decay of $P_G(t)$ 
with the known exponent 0.25. 

\begin{figure}
\noindent \includegraphics[clip,width= 6cm, angle = 270]{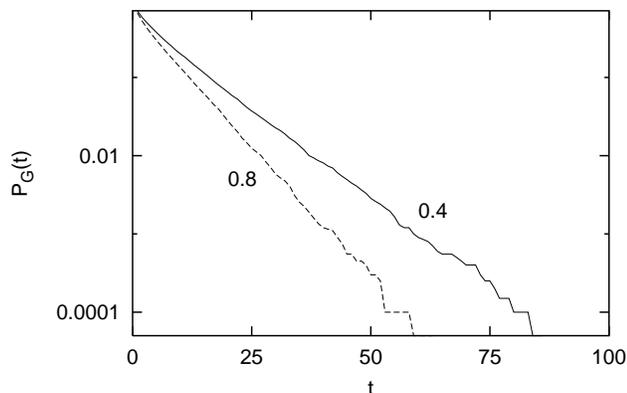}
\caption{The variation of  $P_G(t)$ for $\kappa < 1$ are shown. 
The labels indicate the 
values of $\kappa$.
 The exponential decay is accompanied by different exponents for 
 $0 < \kappa < 0.5$ and $0.5 < \kappa < 1.0$. The system size is $L=12000$ here.}
 \end{figure}
For $0.5 < \kappa < 1$, $P_G(t)$ again has an exponential decay 
\begin{equation}
P_G(t) \sim \exp (-0.16 t). 
\end{equation}
which is faster than that observed in the region  $0 < \kappa < 0.5$ (Eq. (7)).
The results for global persistence for $\kappa < 1$ are shown in Fig. 5.

For $\kappa > 1$ there is algebraic decay of the global persistence (Fig. 2)
\begin{equation}
P_G(t) \sim t^{-\theta^{\prime}_G},
\end{equation}
with $\theta^{\prime}_G \simeq 0.5$, which is different from the exponent 
for the nearest-neighbour Ising model in one dimension. It should be noted that
for $\kappa=0$ the global persistence behaviour could be analysed exactly 
\cite{satya_globe} because the dynamics is essentially that of some 
independent random walkers which annihilate each other. There is no such 
simplification here, as the $j$-th spin in the $\alpha$-th sublattice, namely
$S_{\alpha + 4j}$ is independent of the neighbouring spins in the same
sublattice, but depends on the $j$-th spin in the other sublattices.

\section{Dynamics at $\kappa =1$}

The point $\kappa=1$ signifies a dynamical transition point
with different dynamical behaviour on its two sides and the
behaviour here is different from that for
lower or higher $\kappa$ values. The fraction of persistent spins $P(t)$
decays as a stretched exponential (Fig. 6)
\begin{equation}
P(t) \sim \exp ({-1.92t^{0.21}})
\end{equation}
but this decay is slower than a similar decay (Eq. (2)) for $ 0 < \kappa < 1$.
As regards the behaviour of domain decay, we have found that (i) the
excitation energy $\Delta E $, (ii) the number of defects $D$ (the sites that 
do not belong to the pattern ++ - -), (iii) the number of domains ($M'$)
which are not of length 2, all decay as $t^{-1/z''}$ 
and the lifetime $\tau$ also grows with $L$ as
$L^{z''}$ with $z ''\simeq 3$. 

It is interesting to note that the persistence $P(t)$ does not decay
algebraically which one would 
expect from the fact that the dynamics does lead to the equilibrium state
although at a rate much slower than  that for $\kappa > 1$ \cite{redner}.
This is again, we believe, a special feature related entirely to the
persistence phenomena.

\begin{figure}
\noindent \includegraphics[clip,width= 6cm, angle = 270]{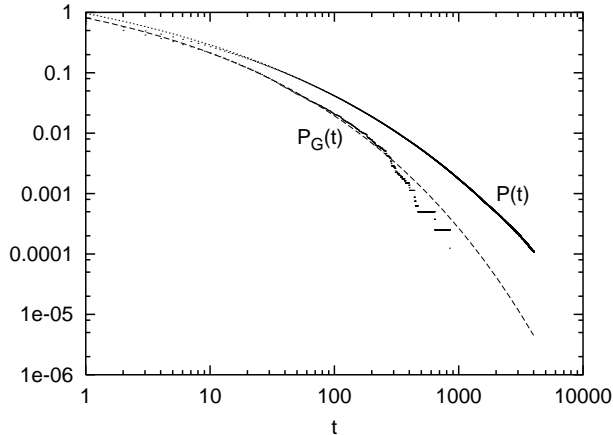}
\caption{The variation of $P(t)$ and $P_G(t)$ for $\kappa =1$ are shown.
Both have a stretched exponential decay, the best fit curves shown in the 
figure are $4.39\exp (-1.69t^{0.25})$ for the global curve and
$6.52\exp (-1.92t^{0.21})$ for the local curve. Data correspond to a $L=8000$ 
system.}
\end{figure}

The dynamical transition point $\kappa = 1$ is marked by a stretched 
exponential decay of the global persistence also 
\begin{equation}
P_G(t) \sim \exp (-1.69 t^{0.25}),
\end{equation}
as shown in Fig. 6.

\section{Summary and conclusions}

In summary, we have studied the persistence behaviour for the ANNNI chain for 
all possible values of $\kappa$, the paramater governing frustration in the 
model.  Three regions of  different behaviour of local persistence are obtained as $\kappa $ is varied. 
The fraction of persistent sites shows a stretched exponential
decay for $\kappa < 1$. At $\kappa = 1$, it is also stretched exponential with
a different exponent.
For $\kappa > 1$, algebraic decay of the persistence
probability is observed with  exponents different from that
of the unfrustrated case. An estimate of the 
dynamical exponent $z^\prime$  using several  approaches  has been made
from which we conclude $z^\prime \simeq 2.30$. This confirms that
the ANNNI model belongs to a different universality class from that of the 
unfrustrated nearest neighbour Ising model for which the corresponding exponent
has a value 2.0.
When global persistence is considered, we again observe different
types of behaviour of the persistent probability in the above three regions.
The behaviour of both local and global persistence
is unique at $\kappa=1$, the  dynamical transition point. 
One should note that there is thus some abrupt change in behaviour (both of
local and of global persistence) at the points $\kappa=0$, $\kappa=1$ and
$1/\kappa=0$. In other words, the behaviour changes sharply as soon as a 
slight amount of next 
nearest-neighbour interaction is added to nearest-neighbour interaction, or
a slight amount of nearest-neighbour interaction is added to next
nearest-neighbour interaction.

It should be mentioned that three regimes of persistence (namely, exponential,
stretched-exponential, and algebraic) have also been studied earlier, although
in somewhat different contexts \cite{satya}.
It should also be pointed out here that all the dynamical features obtained 
here are for the single spin-flip Glauber dynamics. Other types of dynamics, as 
considered in \cite{barma,redner} could lead to different behaviour.

Before we conclude we must also mention that for serial updating (where the sites are
visited serially and updated) the results presented here remains
qualitatively the same in the sense that the indices remain the same upto
the accuracy of the simulation.
Also, for $\kappa > 1$, one obtains the same values of the  exponents even if one starts with a ferromagnetic
state instead of a random state.

Acknowledgments: The authors are grateful to A. Dutta for suggesting the 
problem and thank P. Ray for illuminating discussions. PS acknowledges DST grant number SP/S2/M-11/99.

\end{multicols}
\end{document}